\documentclass[12pt,preprint]{aastex}

\def\Msun{\mbox{\,M$_{\odot}$}}

\newcommand{\ml}{~\Msun ~\rm yr^{-1}}

\def\degs{\ifmmode ^{\circ}\else$^{\circ}$\fi}
\def\amin{\ifmmode ^{\prime}\else$^{\prime}$\fi}
\def\asec{\ifmmode ^{\prime\prime}\else$^{\prime\prime}$\fi}
\def\fdg{\hbox{$.\!\!^\circ$}}          
\def\farcs{\hbox{$.\!\!^{\prime\prime}$}}  
\def\h{\hbox{$^{\rm h}$}}
\def\m{\hbox{$^{\rm m}$}}

\def\degs{\ifmmode ^{\circ}\else$^{\circ}$\fi}
\def\amin{\ifmmode ^{\prime}\else$^{\prime}$\fi}

\def\EE#1{\times 10^{#1}}

\def\cm{\mbox{\,cm}}

\def\cm3{\mbox{\,cm$^{-3}$}}
\def\kms{\mbox{\,km~s$^{-1}$}}

\def\ergshz{\mbox{~erg~s$^{-1}$~Hz$^{-1}$}}
\def\kms{\mbox{\,km s$^{-1}$}}

\def\muJ{\mbox{$\mu$Jy}}
\def\lsim{\!\!\!\phantom{\le}\smash{\buildrel{}\over
 {\lower2.5dd\hbox{$\buildrel{\lower2dd\hbox{$\displaystyle<$}}\over
                                 \sim$}}}\,\,}
\def\gsim{\!\!\!\phantom{\ge}\smash{\buildrel{}\over
{\lower2.5dd\hbox{$\buildrel{\lower2dd\hbox{$\displaystyle>$}}\over
                               \sim$}}}\,\,}

\shorttitle{Radio Detection of SN 2004ip in IRAS 18293-3413}
\shortauthors{P\'erez-Torres et al.}

\begin{document}


\title{Radio Detection of Supernova 2004ip in the Circumnuclear Region
of the Luminous Infrared Galaxy IRAS 18293-3413}


\author{M.A. P\'erez-Torres}
\affil{Instituto de Astrof\'{\i}sica de Andalucia, IAA-CSIC, Apdo. 3004, 18080 Granada, Spain, torres@iaa.es}
\author{S. Mattila\altaffilmark{1}}
\affil{Tuorla Observatory, University of Turku, V\"ais\"al\"antie 20, FI-21500 Piikki\"o, Finland, S.Mattila@qub.ac.uk}
\author{A. Alberdi}
\affil{Instituto de Astrof\'{\i}sica de Andalucia, IAA-CSIC, Apdo. 3004, 18080 Granada, Spain, antxon@iaa.es}
\author{L. Colina}
\affil{Instituto de Estructura de la Materia, IEM-CSIC,  28006 Madrid, Spain, colina@damir.iem.csic.es}
\author{J.M. Torrelles}
\affil{Instituto de Ciencias del Espacio (CSIC)-IEEC, Facultat de F\'{\i}sica,
Universitat de Barcelona, Mart\'{\i} i Franqu\`es 1, 08028 Barcelona, Spain, \\torrelles@ieec.fcr.es}
\author{P. V\"ais\"anen}
\affil{South African Astronomical Observatory, PO Box 9, Observatory 7935, South Africa, petri@saao.ac.za }
\author{S. Ryder}
\affil{Anglo-Australian Observatory, PO Box 296, Epping, NSW 1710, Australia, \\sdr@aao.gov.au}
\author{N. Panagia\altaffilmark{2,3}} 
\affil{STScI, Baltimore, MD 21218, USA, \\panagia@stsci.edu}
\author{A. Wilson}
\affil{Astronomy Department, Univ. of Maryland, College Park, MD 20742, USA, \\wilson@astro.umd.edu}

\altaffiltext{1}{Astrophysics Research Centre, School of Mathematics and Physics, Queen's University Belfast, BT7 1NN, UK.}
\altaffiltext{2}{INAF-CT, Osservatorio Astrofisico di Catania, I-95123 Catania, Italy.}
\altaffiltext{3}{Supernova Ltd. OYV \# 131, Virgin Gorda, British Virgin Islands.}


\begin{abstract}
We report a radio detection of supernova SN 2004ip in the
circumnuclear region of the luminous infrared galaxy \object{IRAS
18293-3413}, using Very Large Array (VLA) observations at 8.4 GHz on
11 June 2007. SN 2004ip had been previously discovered at
near-infrared wavelengths using adaptive optics observations, but its
nature (core-collapse or thermonuclear) could not be definitely
established.  Our radio detection, about three years after the
explosion of the supernova, indicates a prominent interaction of the
ejecta of SN 2004ip with the circumstellar medium, confirming that the
supernova was a core-collapse event (probably a Type II), and strongly
suggesting that its progenitor was a massive star with a significant
mass-loss prior to its explosion.  SN 2004ip has a 8.4 GHz luminosity
of 3.5$\EE{27}$\ergshz, about three times as bright as \object{SN 2000ft} in
NGC 7469 at a similar age, and given its projected distance to the
nucleus ($\sim$500 pc), is one of the closest of all known radio SNe
to a galaxy nucleus, and one of the brightest radio SNe ever.

\end{abstract}

\keywords { --- galaxies: starburst --- 
supernovae: individual (\object{SN 2004ip}) 
--- radio continuum: stars --- radio continuum: galaxies}

\section{Introduction}

Stars more massive than $\sim$8 M$_{\odot}$~explode as 
core-collapse supernovae (CCSNe), i.e., Types Ib/c and II.
Given a reasonable assumption for the initial
mass function (IMF), the observed CCSN rate can be used as 
a direct measure of the current star formation
rate (SFR). Thus, CCSN searches (e.g. Cappellaro et al. 
1999; Dahlen et al. 2004; Cappellaro et al. 2005) can yield a 
new independent measurement of the star formation history
of the Universe. However, optical searches are only able to
discover the SNe not severely affected by dust extinction
and therefore most of the SNe occurring in dusty starburst
galaxies have remained undiscovered even in the local
Universe. 

Furthermore, a large fraction of the star formation at high-$z$ took
place in luminous (L$_{{\rm IR}}$ $>$ 10$^{11}$ L$_{\odot}$; LIRGs)
and ultraluminous (L$_{{\rm IR}}$ $>$ 10$^{12}$ L$_{\odot}$; ULIRGs)
infrared galaxies (P\'erez-Gonz\'alez et al. 2005) where the dust
extinction is even more severe.  The existence of hidden SN factories
in the nuclear and circumnuclear regions of (U)LIRGs has already been
demonstrated by high-resolution radio observations. For example, VLA
observations of the circumnuclear starburst in NGC 7469 revealed an
extremely bright radio supernova, SN 2000ft (Colina et
al. 2001b). More recently, very long baseline interferometry (VLBI)
observations of the nearby ULIRG, Arp 220, have revealed luminous
radio SNe within its innermost 150 pc nuclear regions at a rate of
4$\pm$2 yr$^{-1}$, which indicates a SFR high enough to power its
entire IR luminosity (Lonsdale et al. 2006). These studies confirm
that CCSN rates a couple of orders of magnitude higher than in
ordinary field galaxies can be expected for starburst dominated
(U)LIRGs.

Ground-based near-infrared (IR) observations (with spatial resolution
$\sim$1 arcsec) can provide a complementary tool to high-resolution
radio observations in the search for CCSNe in (U)LIRGs (Van Buren et
al. 1994; Grossan et al. 1999; Mattila \& Meikle 2001; Maiolino et
al. 2002; Mannucci et al. 2003; Mattila et al. 2004; Mattila et
al. 2005a,b).  More recently, Mattila et al. (2007a,b) reported the
first-ever adaptive optics (AO) assisted discovery of a SN making use
of the NAOS CONICA (NACO) AO system on the ESO Very Large Telescope
(VLT). As a result of their K$_{\rm S}$-band [2.2 microns] search for
highly-obscured CCSNe in a sample of (U)LIRGs (see also V\"ais\"anen
et al. 2007), SN 2004ip was discovered in IRAS 18293-3413, a
LIRG at a distance of 79 Mpc (at this distance 1 arcsecond corresponds
to 383 pc), with an estimated host galaxy extinction towards SN 2004ip
between 5 and 40 magnitudes in $A_{\rm V}$
(Mattila et al. 2007b). SN 2004ip is located in the nuclear regions of IRAS
18293-3413 at 1".14 east and 0".78 north of (about 500 pc projected
distance from) the galaxy's K$_{\rm s}$-band nucleus.  

It has been proposed that CCSNe exploding in dense environments, like
those encountered in the circumnuclear regions of starburst
galaxies, can result in very luminous radio SNe (Chevalier 1982,
Chugai 1997). An apparent confirmation of such a theory came from the
discovery of the radio bright SN 2000ft (Colina et al. 2001a,b) in a
circumnuclear starburst ring ($\sim$600 pc from the nucleus) of its
LIRG host NGC 7469 ($D\approx$~70~Mpc).  However, the case of SN
2000ft is outstanding because its six-year long VLA monitoring has
shown that the supernova shares essentially the same properties that
are common to radio SNe identified as Type II SNe, despite having
exploded in the dusty and very dense environment of the circumnuclear
region (Alberdi et al. 2006). We also note that the optical glow of SN 2000ft
has recently been identified in archival HST images taken on 13 May 2000
(Colina et al. 2007), consistent with the predicted date of the
explosion from the analysis of its radio light curves (Alberdi et
al. 2006).

The observations of SN 2004ip at near-IR wavelengths did not allow it to
be unambiguously identified with either a core-collapse (Type Ib/c, or II SN),
or thermonuclear (Type Ia) event.  Furthermore, the discovery of SN 2004ip
could only be announced very recently (Mattila et al. 2007a), thus
making any follow-up observations difficult. However, assuming that SN
2004ip was a \object{SN 2000ft}-like event, we could expect it to be bright
enough as to be detectable with the VLA even three years after its
explosion. Hence, we proposed rapid response science VLA time to
search for cm-wavelength radio emission at its location. In this
Letter, we report the first radio detection of SN 2004ip, indicating a
strong interaction with its circumstellar medium.

\section{VLA Observations}

We observed the host galaxy of SN 2004ip, IRAS 18293-3413, on 11 June
2007 at the frequency of 8.4 GHz with the VLA in A configuration,
aimed at resolving and detecting SN 2004ip at radio wavelengths.  We
used rapid response exploratory VLA Time, since the angular distance
of the supernova to the nucleus (about 1.4 arcsec) made it necessary
to ask for the VLA in its most extended, A-configuration.  We observed
at 8.4 GHz, which resulted in an angular resolution of 0\farcs62 and
0\farcs21 in right ascension (R.A.) and declination (decl.),
respectively, enough to discern the radio emission from the supernova
from that of the nucleus. Also, at 8.4 GHz any extended radio
emission from the galaxy and its nucleus should be less prominent than
at lower frequencies (the only previously existing VLA radio
observations of IRAS 18293-3413 were carried-out at 1.4 GHz with an
angular resolution of about 5 arcseconds, indicating a large flux
density, $\sim$130 mJy for the entire galaxy), while at higher
frequencies the supernova was expected to be too faint to be
detectable.

The observations lasted for two hours, and consisted of $\sim$11.5~min
scans on SN 2004ip (for a total on-target time of $\sim$81 min),
interleaved with $\sim$2.5~min scans on the phase and amplitude
calibrator J1820-254, and each time ending with a $\sim$5~min
observation of the quasar 3C~286 (1331+305) to set the absolute flux
density scale.  We edited, calibrated, and imaged our 8.4 GHz VLA data
by following standard data reduction techniques implemented within the
NRAO Astronomical Image Processing System ({\it AIPS}).

\section{Results}

Our results are summarized in Figure 1 and Table 1.  Figure 1 shows
the 8.4 GHz radio emission from the central parts of IRAS 18293-3413
(contours) overlaid on a subtracted NACO K$_{\rm S}$-band image (shown
with an inverted brightness scale), which is the result of subtracting
a K$_{\rm S}$-band image obtained on 4th May 2004 from the image
obtained on 15th September 2004 (for details see Mattila et al. 2007b). 
A strong negative residual coincident
with the K$_{\rm S}$-band nucleus is visible in white color, and SN
2004ip is clearly visible as a positive point source (in black
colour).  A local maximum of radio emission within the circumnuclear
region of the galaxy is right coincident with SN~2004ip. 

The radio contours also indicate evidence for a number of bright,
compact objects whose discussion is beyond the scope of this
Letter. We only note here that we have found eight such compact
regions with signal-to-noise ratios larger than 20 ($S_\nu \ga
460$\muJ), but which have no clear counterpart in the near-IR image.
Future multi-wavelength VLA observations of this galaxy will allow us to
shed some light on the nature of these compact sources.  In Table 1, we show
the 8.4 GHz VLA flux densities of the nucleus of IRAS 18293-3413 
and a source we identify as SN 2004ip, which exploded
between 4th May and 13th September 2004 within the nuclear starburst
of the galaxy.

The absolute position of SN~2004ip was only reported with an estimated
precision of $\pm$0.4'' by Mattila et al. (2007b). To compare with the
position of the source detected in our VLA image (see Fig. 1), the
precision of the supernova astrometry first needed to be improved. For
this purpose we used archival K-band data of IRAS 18293-3413 obtained
with the SOFI near-IR camera on the New Technology Telescope (NTT) on
8th September 2001. The final K-band image, reduced using standard
IRAF routines, has an on-source exposure time of 30 min and a seeing
FWHM of $\sim$1.2''. Due to the low galactic latitude a large number
of stars is visible within the $\sim$7'' $\times$ 7'' field of view of
the combined NTT image. Therefore, we were able to identify over 400
bright and isolated stars with astrometry available from 2MASS. These
yielded a world coordinate system (WCS) solution for the NTT image
with rms of 4 and 3 milliarcsec (mas) in R.A. and Decl.,
respectively. We then used the centroid coordinates of 14 isolated
stars, detected within the field of view of both the 42'' $\times$
42'' NACO and NTT images, to align the images in IRAF using a general
geometric transformation with no non-linear part. This yielded an rms of
34 and 22 mas in R.A. and Decl., respectively. Finally, the position
of the supernova was measured in a NACO K$_{\rm S}$-band image which
is the result of subtracting an image obtained on 4th May 2004 from an
image obtained on 15th September 2004 (for details on the images and
the subtraction method see Mattila et al. 2007b).  We adopted the
average from three different methods (centroid, gauss, ofilter) used
in IRAF as the SN position. The position uncertainty as indicated by
the standard deviation of these measurements was very small $\sim$1
mas for both $x$ and $y$, thanks to the well sampled NACO PSF and the
flat and close to zero background in the subtracted image. Using the
WCS of the NTT image aligned to the NACO image, we obtained R.A. =
18h32m41.207s and Decl. = -34$^{o}$11'26.80'' (equinox 2000.0) for the
SN with an estimated precision of $\pm$34 and $\pm$22 mas in R.A. and
Decl., respectively.

Therefore, the radio source we identified as SN 2004ip appears
coincident with the near-IR position of the supernova within 10 and 20
mas in R.A. and Decl., respectively. This is within the uncertainties
of the SN near-IR position (as derived above) and radio position
($\pm$20 mas and $\pm$10 mas in R.A. and Decl., respectively), which
confirms that the radio emission corresponds to SN 2004ip.

\section{Discussion and Summary}

Thermonuclear (Type Ia) supernovae are not expected to be strong radio
emitters, and have not yet been detected at radio wavelengths (e.g.,
Panagia et al. 2006).  Current modelling of their radio emission
indicates that the circumstellar wind around the progenitor star is
much less dense than in the case of CCSNe, and would be overrun in
about one day due to its proximity and the much higher velocity of the
supernova blast wave.  At the distance of IRAS 18293-3413, this would
result in an SN Ia radio emission reaching 8.4 GHz peak values of
(13--50)\,$\mu$Jy at around (3--10)\,days after the explosion, and quickly
decreasing below (0.3--3)\,$\mu$Jy after a hundred days, depending on the
particular model (P. Lundqvist, private communication).  Therefore, a
thermonuclear origin for SN 2004ip --detected at $\sim 470 \mu$Jy
after more than 1000 days after its explosion-- can be ruled out.

CCSNe are expected --as opposed to thermonuclear SNe-- to become
strong radio emitters when the SN ejecta interact with the
circumstellar medium (CSM) that was ejected by the progenitor star
before its explosion as a supernova (Chevalier 1982; Weiler et
al. 1986).  Indeed, the interaction gives rise to a high-energy
density shell, which is Rayleigh-Taylor unstable and can drive
turbulent motions that may amplify the existing magnetic field and
efficiently accelerate relativistic electrons, thus enhancing the
emission of synchrotron radiation at radio wavelengths (Chevalier
1982).  The duration of this radio SN phase is limited, however, by
the extent of the expanding wind of the progenitor star, which can reach
a radius where the ram pressure of the wind, $\rho_w\,v_w^{2}$,
equals the external pressure of the insterstellar medium ISM,
$P_{ISM}$ (Chevalier \& Fransson 2001).  For a spherically symmetric,
steady wind ($\rho_w \propto r^{-2}$), this radius is $r_w \approx
0.18\,\dot M_{-4}^{1/2}v_{w1}^{1/2}p_7^{-1/2}$ pc, where $\dot M_{-4}$
is the mass loss rate in units of $10^{-4}\ml$, $v_{w1}$ is the wind
velocity in units of $10\kms$, and $p_7$ is the ISM pressure in units
of $10^7$ cm$^{-3}$ K, which is the estimated pressure for the central
region of the starburst in M~82 (Chevalier \& Clegg 1985). At this
distance, a CCSN would then enter the supernova remnant (SNR) phase,
and the radio emission would no longer be due to the interaction with
the CSM, but rather with the ISM. However, even if the ejecta of SN
2004ip would have been freely expanding at a constant velocity of $v_s
= 10^4$\kms, the distance reached by the circumstellar shock in three
years would be $r \lsim 0.03$ pc, which is much smaller than $r_w$
assuming an ISM pressure similar to M~82.  
We conclude that SN 2004ip
is still in its SN phase and its radio emission is being powered
--even three years after the SN explosion-- by prominent interaction
with the CSM. This fact confirms that the supernova was a
core-collapse event, and implies that its progenitor was a massive
star with a significant mass-loss ($\dot M \ga 10^{-4}\ml$) prior to
its core collapse.

SN 2004ip exploded sometime between 4th May and 13th September 2004,
so its radio emission is now likely in a decaying phase.  The 8.4 GHz
flux density of SN 2004ip detected in the June 2007 observations
($\sim$470$\mu$Jy) correspond to an isotropic luminosity of
3.5$\EE{27}$\ergshz, or three times as luminous as \object{SN 2000ft}
in NGC 7469 at a similar age.  Given the fact that it exploded about
three years earlier, SN 2004ip might have been one of the most
luminous radio SNe ever at its peak (and probably significantly
brighter than the nucleus of its host galaxy), and would therefore
belong to the class of extremely bright and long-lasting radio SNe,
like SN 1978K (Schlegel et al. 1999), \object{SN 1986J} (Weiler et
al. 1990; P\'erez-Torres et al. 2002), SN 1988Z (Van Dyk et
al. 1993b), or SN 2000ft (Colina et al. 2001b), which at their peak
emission were a thousand to a few thousand times more luminous than
Cas A, the brightest radio SN in the Milky Way (Weiler et al. 1986).
In addition to its extreme radio brightness, SN 2004ip shares with
\object{SN 2000ft} in the LIRG NGC 7469 (Alberdi et al. 2006), and
with the young radio supernovae in the central regions of Arp 299
(Neff et al. 2004), other characteristics in common.  Indeed, SN
2004ip exploded in the circumnuclear region of a Luminous Infrared
Galaxy, is located at a similar projected distance ($r\sim 500$
pc) from the galaxy nucleus, is still in its radio SN phase, and is
detectable several years after its explosion, These facts indicate
that such radio SNe might be a relatively common phenomenon in
circumnuclear starburst environments.

Extremely bright and long lasting radio SNe are identified in the
optical as Type II supernovae, and because of their huge radio
luminosities, their progenitors are believed to be massive stars in
the 20 to 30 \Msun~ range (Weiler et al. 1990; Van Dyk et al. 1993)
that explode in very dense environments (e. g., Chugai 1997). The
typical spectral index for normal Type II supernovae is $-$0.6 to
$-$0.8 (Weiler et al. 2002). Type II radio SNe also exhibit slow rises
and declines, and normally take more than one year to reach their 8.4
GHz peak radio emission.  On the other hand, Type Ib/c supernovae can
also produce bright radio SNe, e.g., SN 1983N (Sramek et al. 1984;
Weiler et al. 1986) and SN 1990B (van Dyk et al.  1993a), and the
brightest of all, SN 1998bw (Kulkarni et al. 1998).  However, the
radio emission of these supernovae is characterized by a rather fast
rise and decay (few days to a few weeks), and by radio spectral
indices of about $-$1.1 to $-$1.2 (Weiler et al. 2002).  While the 
measured single radio flux of SN 2004ip does not allow a
definite classification, the fact that its radio luminosity is so high
even three years after the explosion favours strongly a Type II origin
since their radio flux densities tend to decrease much more slowly
than the radio emission from Type Ib/c SNe.

The level of the 8.4 GHz radio emission of SN 2004ip detected in our
June 2007 observation ($\sim$470$\mu$Jy) suggests that we might be
able to monitor the radio flux density evolution of SN 2004ip with the
VLA for some years from now, thus probing the circumstellar
interaction around the supernova, eventually allowing us to detect its
transition from the SN phase to the SNR phase, as has been suggested
for a number of compact sources in the nuclear starburst of Arp 220
(Parra et al. 2007). In addition, a radio light curve follow-up could
provide more detailed information on the progenitor of SN 2004ip. For
example, Ryder et al. (2004) inferred the action of a binary companion
from periodic modulations in the radio light curve of SN 2000ig, which
was subsequently confirmed from imaging at Gemini (Ryder et al. 2006).
This would make SN 2004ip the second case, only after \object{SN
2000ft} in NGC 7469, where such radio monitoring has been carried out
for a SN in the circumnuclear starburst of a LIRG, and will help to
better understand the behaviour of SNe within dense starburst
environments (e.g., Alberdi et al. 2006).  Furthermore, Mattila et
al. (2007b) estimated an average SFR of 135 M$_{\odot}$~yr$^{-1}$ for
IRAS 18293-3413 corresponding to a CCSN rate of about 1.0 SN per
year. Therefore, radio monitoring at high-resolution and sensitivity
with the VLA might easily result also in the discovery of new radio
SNe.

The number of CCSNe discovered in circumnuclear regions of (U)LIRGs,
both at infrared and radio wavelengths, is still small. However, these
events may have an important impact when estimating the {\it complete}
local CCSN rates including also the SNe in the optically obscured
parts of the galaxies. The direct detection and study of CCSNe in
starburst galaxies over a large range of IR luminosities (L$_{\rm IR}$
= 10$^{10}$--10$^{12}$ L$_{\odot}$) is also crucial for interpreting
the results of the high-z CCSN searches, since a large fraction of the
massive star formation at high-$z$ took place in IR luminous galaxies.
Therefore, our discovery shows that the combination of high-resolution
observations at near-infrared and radio wavelengths is a powerful tool
to search for CCSN events from LIRGs in the local Universe, and thus
establish their core-collapse supernova rates and star formation
rates, as well as to constrain the nature of the discovered events
through their interaction with the surrounding medium.

\acknowledgments 
We are grateful to an anonymous referee for helpful comments on the
manuscript.  We are also grateful to the National Radio Astronomy
Observatory (NRAO) for granting us Rapid Response, Very Large Array
(VLA) time for this project. NRAO is a facility of the National
Science Foundation operated under cooperative agreement by Associated
Universities Inc.  MAPT research is supported by the Ram\'on y Cajal
programme of the Spanish Spanish Ministry of Education and Science.
MAPT and AA, and JMT acknowledge support from the Spanish grants
AYA2006-14986-C02-C01 and AYA2005-08523-C03, respectively.  SM
acknowledges financial support from the Participating Organisations of
EURYI and the EC Sixth Framework Programme and from the Academy of
Finland (project: 8120503).

\clearpage

\clearpage

\begin{deluxetable}{lccr}
\tablecaption{8.4 GHz VLA positions and flux densities
      of SN 2004ip and the nucleus of its host galaxy IRAS 18293-3413}
\tablehead{
\colhead{Component} & \colhead{$\alpha$(J2000)} &  \colhead{$\delta$(J2000)} & 
\colhead{S$_{\nu}$ [$\mu$Jy]}
} 
\startdata
     SN 2004ip  & 18\h 32\m 41\fs208 & -34\degs11\amin26\farcs82 
                                                    &  466 $\pm$ 25 \\
     Nucleus    & 18\h 32\m 41\fs108 & -34\degs11\amin27\farcs62 
                                                    & 1708 $\pm$ 41 \\
\enddata
\tablecomments{The errors in the flux density measurements
represent one statistical standard
deviation, $\sigma_f$, and are calculated as 
$\sigma_f^2 = (\epsilon\,S_0)^2 + \sigma_0^2$, where $S_0$ is the
measured peak flux density, $\sigma_0$ is the off-source rms, and the
fractional error $\epsilon$
(assumed to be of 2\%) accounts for the inaccuracy
of VLA flux density calibration and possible deviations of the primary
calibrator from an absolute flux density scale.
}
\end{deluxetable}

\clearpage

\begin{figure}[tbh]
\epsscale{1.0}
\plotone{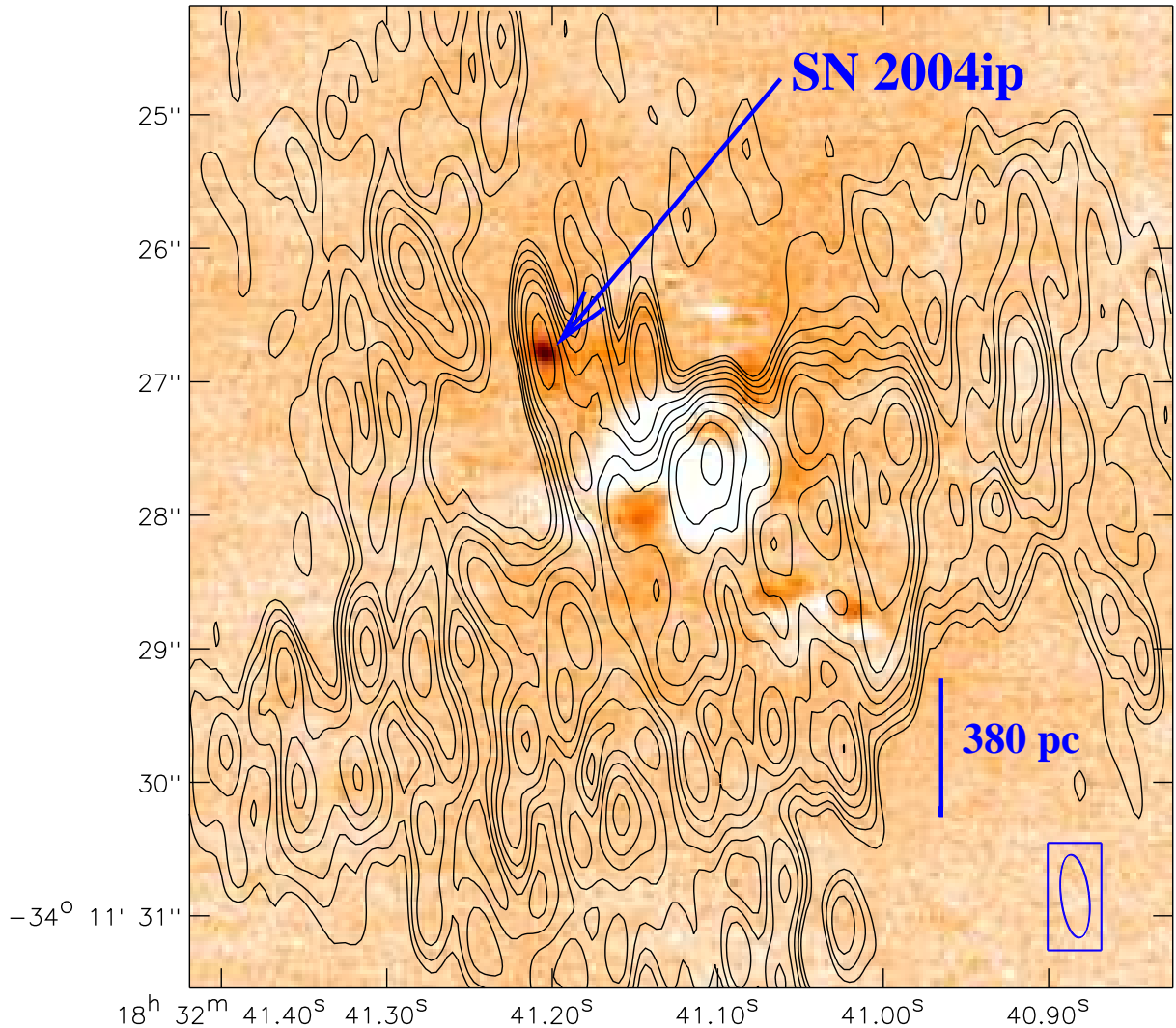}
\caption{Contours of 8.4~GHz observations of IRAS~18293-3413 made on
  11 June 2007 with the VLA in A configuration, overlaid on a NACO
  K$_{\rm S}$-band (near-IR) image (shown with an inverted brightness
  scale).  A local maximum of radio emission within the circumnuclear
  region of the galaxy is coincident with SN~2004ip.  The VLA contours
  are drawn at (-2,2,$2\,\sqrt{2}$,4,...)$\times$ the off-source rms
  noise in the image, which is 23\,$\mu$Jy beam$^{-1}$, and the
  synthesized beam size is of 0\farcs62 $\times$ 0\farcs21 along
  P.A. 6\fdg5.  The 8.4 GHz peak brightness is 1.71 mJy beam$^{-1}$,
  and corresponds to the radio nucleus of IRAS 18293-3413.  The total
  continuum 8.4 GHz emission from IRAS 18293-3413 above 3\,$\sigma$ is
  40.1 mJy.}
\end{figure}

\end{document}